\documentclass[a4paper]{PoS}
\usepackage{amssymb}
\usepackage{amsmath}
\usepackage[english]{babel}
\usepackage{amsfonts}
\usepackage{cite}
\usepackage[utf8]{inputenc}

\input{tcilatex}
\title{Fermion mass and mixing pattern in a minimal T7 flavor 331 model}
\ShortTitle{A minimal T7 flavor 331 model}

\author{\speaker{A. E. C\'arcamo Hern\'andez}\thanks{A footnote may follow.}\\
       Universidad T\'{e}cnica Federico Santa Mar\'{\i}a\\
and Centro Cient\'{\i}fico-Tecnol\'{o}gico de Valpara\'{\i}so\\
Casilla 110-V, Valpara\'{\i}so, Chile,\\
        E-mail: \email{antonio.carcamo@usm.cl}}

\author{R. Martinez\\
        Universidad Nacional de Colombia, Departamento de F\'{\i}sica,
Ciudad Universitaria, Bogot\'{a} D.C., Colombia.
        E-mail: \email{remartinezm@unal.edu.co}}

\abstract{We present a model based on the $SU(3)_{C}\otimes SU(3)_{L}\otimes U(1)_{X}$ gauge symmetry having an extra $T_{7}\otimes Z_{2}\otimes Z_{3}\otimes Z_{14} $ flavor group, which successfully describes the observed SM fermion mass and mixing pattern. In this framework, the light active neutrino masses arise via double seesaw mechanism and the observed charged fermion mass and quark mixing hierarchy is a consequence of the $Z_{2}\otimes Z_{3}\otimes Z_{14}$ symmetry breaking at very high energy. In our minimal $T_{7}$ flavor 331 model, the spectrum of neutrinos includes very light active neutrinos as well as heavy and very heavy sterile neutrinos. The obtained physical observables for both quark and lepton sectors are compatible with their experimental values. The model predicts the absence of leptonic Dirac CP violating phase.}

\FullConference{18th International Conference From the Planck Scale to the Electroweak Scale \\
		25-29 May 2015\\
		Ioannina, Greece }

\begin{document}

\section{Introduction}

Despite its big experimental success, the Standard Model (SM) has several
unexplained features. Some of them are the Dark Matter problem, the fermion
mass and mixing hierarchy and the neutrino oscillations. The lack of
predictivity of the Standard Model Yukawa sector, motivates to consider
extensions of the Standard Model aimed to address its flavor puzzle.
Discrete flavor symmetries are important because they generate fermion
textures useful to explain the three generation flavor structure. Very recently,
discrete groups such as, for example $A_{4}$ \cite{Hernandez:2013dta,Hernandez:2015tna,Campos:2014lla}, $S_4$ \cite{Campos:2014zaa}, $S_{3}$ 
\cite{Hernandez:2013hea,Hernandez:2014vta,Hernandez:2014lpa,Hernandez:2015dga,Hernandez:2015zeh}, $T_{7}$ \cite{Hernandez:2015cra,Arbelaez:2015toa} and so forth, 
have been implemented in several extensions of the SM, to explain the observed pattern of fermion masses and mixings.


\quad Furthermore, another unaswered issue in particle physics is the
existence of three generations of fermions at low energies. The mixing
patterns of leptons and quarks are significantly different; while in the
quark sector, the mixing angles are small, in the lepton sector two of the
mixing angles are large and one is small. Models having $SU(3)_{C}\otimes
SU(3)_{L}\otimes U(1)_{X}$ as a gauge symmetry, are vectorlike with three
fermion generations and thus do not contain anomalies \cite{CarcamoHernandez:2005ka,Hernandez:2013mcf}. Defining the electric charge as
the linear combination of the $T_{3}$ and $T_{8}$ $SU(3)_{L}$ generators, we
have that it is a free parameter, which does not depend on the anomalies ($%
\beta $). The charge of the exotic particles is defined by setting a value
for the $\beta $ parameter. Setting $\beta =-\frac{1}{\sqrt{3}}$, implies
that the third component of the weak lepton triplet is a neutral field $\nu
_{R}^{C}$ allowing to build the Dirac Yukawa term with the usual field $\nu
_{L}$ of the weak doublet. The 331 models with $\beta =-\frac{1}{\sqrt{3}}$
provide an alternative neutrino mass generation mechanism and include in
their neutrino spectrum light active sub-eV scale neutrinos as well as
sterile neutrinos which could be dark matter candidates if they are light
enough or candidates for detection at the LHC, if they have TeV scale
masses. Having TeV scale sterile neutrinos in its neutrino spectrum, makes
the 331 models very important since if these sterile neutrinos are detected
at the LHC, these models can shed light in the understanding of the
electroweak symmetry breaking mechanism.

\quad In this paper we formulate an extension of the minimal $%
SU(3)_{C}\times SU(3)_{L}\times U(1)_{X}$ model with $\beta =-\frac{1}{\sqrt{%
3}}$, where an extra \mbox{$T_{7}\otimes Z_{2}\otimes Z_{3}\otimes Z_{14}$}
discrete group extends the symmetry of the model and very heavy extra scalar
fields are added with the aim to generate viable and predictive textures for
the fermion sector that successfully describe the fermion mass and mixing
pattern.



\section{The model}
\label{model}

We consider a $SU(3)_{C}\otimes SU(3)_{L}\otimes U(1)_{X}\otimes
T_{7}\otimes Z_{2}\otimes Z_{3}\otimes Z_{14}$ model where the full symmetry 
$\mathcal{G}$ is spontaneously broken in three steps as follows: 
\begin{eqnarray}
&&\mathcal{G}=SU(3)_{C}\otimes SU\left( 3\right) _{L}\otimes U\left(
1\right) _{X}\otimes T_{7}\otimes Z_{2}\otimes Z_{3}\otimes Z_{14}{%
\xrightarrow{\Lambda _{int}}}  \notag \\
&&\hspace{7mm}SU(3)_{C}\otimes SU\left( 3\right) _{L}\otimes U\left(
1\right) _{X}{\xrightarrow{v_{\chi }}}SU(3)_{C}\otimes SU\left( 2\right)
_{L}\otimes U\left( 1\right) _{Y}{\xrightarrow{v_{\eta },v_{\rho }}}  \notag
\\
&&\hspace{7mm}SU(3)_{C}\otimes U\left( 1\right) _{Q},  \label{Group}
\end{eqnarray}%
where the hierarchy $v_{\eta },v_{\rho }\ll v_{\chi }\ll \Lambda _{int}$
among the symmetry breaking scales is fullfilled.


\quad The electric charge in our 331 model is defined as: 
\begin{equation}
Q=T_{3}-\frac{1}{\sqrt{3}}T_{8}+XI,
\end{equation}%
where $T_3$ and $T_8$ are the $SU(3)_L$ diagonal generators and $I$ is the $%
3\times 3$ identity matrix. 

\quad The anomaly cancellation requirement implies that quarks are unified
in the following $(SU(3)_{C},SU(3)_{L},U(1)_{X})$ left- and right-handed
representations: 
\begin{align}
Q_{L}^{1,2}& =%
\begin{pmatrix}
D^{1,2} \\ 
-U^{1,2} \\ 
J^{1,2} \\ 
\end{pmatrix}%
_{L}:(3,3^{\ast },0),\hspace{1cm}Q_{L}^{3}=%
\begin{pmatrix}
U^{3} \\ 
D^{3} \\ 
T \\ 
\end{pmatrix}%
_{L}:(3,3,1/3),  \label{fermion_spectrumleft} \\
& 
\begin{array}{c}
D_{R}^{1,2,3}:(3^{\ast },1,-1/3), \\ 
J_{R}^{1,2}:(3^{\ast },1,-1/3),%
\end{array}%
\hspace{0.7cm}%
\begin{array}{c}
U_{R}^{1,2,3}:(3^{\ast },1,2/3), \\ 
T_{R}:(3^{\ast },1,2/3).%
\end{array}
\label{fermion_spectrumright}
\end{align}%
Here $U_{L}^{i}$ and $D_{L}^{i}$ ($i=1,2,3$) are the left handed up- and
down-type quarks in the flavor basis. The right handed SM\ quarks $U_{R}^{i}$
and $D_{R}^{i}$ ($i=1,2,3$) and right handed exotic quarks $T_{R}$ and $%
J_{R}^{1,2}$ are assigned into $SU(3)_{L}$ singlets representations, so that
their $U(1)_{X}$ quantum numbers correspond to their electric charges.

Furthermore, cancellation of anomalies implies that leptons are grouped in
the following $(SU(3)_{C},SU(3)_{L},U(1)_{X})$ left- and right-handed
representations: 
\begin{align}
L_{L}^{1,2,3}& =%
\begin{pmatrix}
\nu ^{1,2,3} \\ 
e^{1,2,3} \\ 
(\nu ^{1,2,3})^{c} \\ 
\end{pmatrix}%
_{L}:(1,3,-1/3), \\
& 
\begin{array}{c}
e_{R}:(1,1,-1), \\ 
N_{R}^{1}:(1,1,0), \\ 
\end{array}%
\hspace{0.7cm}%
\begin{array}{c}
\mu _{R}:(1,1,-1), \\ 
N_{R}^{2}:(1,1,0), \\ 
\end{array}%
\hspace{0.7cm}%
\begin{array}{c}
\tau _{R}:(1,1,-1), \\ 
N_{R}^{3}:(1,1,0). \\ 
\end{array}%
\end{align}
where $\nu _{L}^{i}$ and $e_{L}^{i}$ ($e_{L},\mu _{L},\tau _{L}$) are the
neutral and charged lepton families, respectively. Let's note that we assign
the right-handed leptons as $SU(3)_{L}$ singlets.
The exotic leptons of the model are: three neutral Majorana leptons $(\nu
^{1,2,3})_{L}^{c}$ and three right-handed Majorana leptons $N_{R}^{1,2,3}$. 

\quad The scalar sector the 331 models includes: three $3$'s irreps of $%
SU(3)_{L}$, where one triplet $\chi $ gets a TeV scale vaccuum expectation
value (VEV) $v_{\chi }$, that breaks the $SU(3)_{L}\times U(1)_{X}$ symmetry
down to $SU(2)_{L}\times U(1)_{Y}$, thus generating the masses of non SM
fermions and non SM gauge bosons; and two light triplets $\eta $ and $\rho $
acquiring electroweak scale VEVs $v_{\eta }$ and $v_{\rho }$, respectively
and thus providing masses for the fermions and gauge bosons of the SM.

\quad Regarding the scalar sector of the minimal 331 model, we assign the
scalar fields in the following $[SU(3)_{L},U(1)_{X}]$ representations: 
\begin{align}
\chi & =%
\begin{pmatrix}
\chi _{1}^{0} \\ 
\chi _{2}^{-} \\ 
\frac{1}{\sqrt{2}}(\upsilon _{\chi }+\xi _{\chi }\pm i\zeta _{\chi }) \\ 
\end{pmatrix}%
:(3,-1/3),\hspace{1cm}\rho =%
\begin{pmatrix}
\rho _{1}^{+} \\ 
\frac{1}{\sqrt{2}}(\upsilon _{\rho }+\xi _{\rho }\pm i\zeta _{\rho }) \\ 
\rho _{3}^{+} \\ 
\end{pmatrix}%
:(3,2/3),  \notag \\
\eta & =%
\begin{pmatrix}
\frac{1}{\sqrt{2}}(\upsilon _{\eta }+\xi _{\eta }\pm i\zeta _{\eta }) \\ 
\eta _{2}^{-} \\ 
\eta _{3}^{0}%
\end{pmatrix}%
:(3,-1/3).  \label{331-scalar}
\end{align}
We extend the scalar sector of the minimal 331 model by adding the following
eleven very heavy $SU(3)_{L}$ scalar singlets: 
\begin{equation}
\sigma \sim (1,0),\hspace{1cm}\tau \sim (1,0),\hspace{0.5cm}\xi _{j}:(1,0),%
\hspace{0.5cm}\zeta _{j}:(1,0),\hspace{0.5cm}S_{j}:(1,0),\hspace{0.5cm}%
j=1,2,3.  \label{331scalarsextra}
\end{equation}
We assign the scalars into $T_{7}$ triplet, $T_{7}$ antitriplet and $T_{7}$
singlet representions. The $T_{7}\otimes Z_{2}\otimes Z_{3}\otimes Z_{14}$
assignments of the scalar fields are: 
\begin{eqnarray}
\eta &\sim &\left( \mathbf{1}_{0},1\mathbf{,}e^{\frac{2\pi i}{3}},1\right) ,%
\hspace{1cm}\rho \sim \left( \mathbf{1}_{0}\mathbf{,}1,e^{-\frac{2\pi i}{3}%
},1\right) ,\hspace{0.5cm}\chi \sim \left( \mathbf{1}_{0}\mathbf{,}%
1,1,1\right) ,\hspace{0.5cm}\tau \sim \left( \mathbf{1}_{1}\mathbf{,-}%
1,1,1\right),  \notag \\
\xi &\sim &\left( \mathbf{3,}1,e^{\frac{2\pi i}{3}},1\right) ,\hspace{0.5cm}%
\zeta \sim \left( \overline{\mathbf{3}}\mathbf{,}1,1,1\right) ,\hspace{0.5cm}%
S\sim \left( \mathbf{3,}1,e^{-\frac{2\pi i}{3}},1\right) ,\hspace{0.5cm}%
\sigma \sim \left( \mathbf{1}_{0}\mathbf{,}1,1,e^{-\frac{i\pi }{7}}\right) .
\end{eqnarray}
In the concerning to the lepton sector, we have the following $%
T_{7}\otimes Z_{2}\otimes Z_{3}\otimes Z_{14}$ assignments: 
\begin{eqnarray}
L_{L} &\sim &\left( \mathbf{3,}1,e^{\frac{2\pi i}{3}},1\right) ,\hspace{1cm}%
e_{R}\sim \left( \mathbf{1}_{0}\mathbf{,}1,e^{\frac{2\pi i}{3}},-1\right) ,%
\hspace{1cm}\mu _{R}\sim \left( \mathbf{1}_{1}\mathbf{,}1,e^{\frac{2\pi i}{3}%
},e^{\frac{4i\pi }{7}}\right) ,  \notag \\
\tau _{R} &\sim &\left( \mathbf{1}_{2}\mathbf{,}1,e^{\frac{2\pi i}{3}},e^{%
\frac{2i\pi }{7}}\right) ,\hspace{1cm}N_{R}\sim \left( \mathbf{3,}1,e^{\frac{%
2\pi i}{3}},1\right) ,
\end{eqnarray}%
while the $T_{7}\otimes Z_{2}\otimes Z_{3}\otimes Z_{14}$ assignments for
the quark sector are: 
\begin{eqnarray}
Q_{L}^{1} &\sim &\left( \mathbf{1}_{0}\mathbf{,}1,1,e^{\frac{2\pi i}{7}%
}\right) ,\hspace{1cm}Q_{L}^{2}\sim \left( \mathbf{1}_{0}\mathbf{,}1,1,e^{%
\frac{\pi i}{7}}\right) ,\hspace{1cm}Q_{L}^{3}\sim \left( \mathbf{1}_{0}%
\mathbf{,}1,1,1\right) ,  \notag \\
U_{R}^{1} &\sim &\left( \mathbf{1}_{0}\mathbf{,}1,e^{-\frac{2\pi i}{3}},e^{%
\frac{2\pi i}{7}}\right) ,\hspace{1cm}U_{R}^{2}\sim \left( \mathbf{1}_{0}%
\mathbf{,}1,e^{-\frac{2\pi i}{3}},e^{\frac{\pi i}{7}}\right) ,\hspace{1cm}%
U_{R}^{3}\sim \left( \mathbf{1}_{0}\mathbf{,}1,e^{-\frac{2\pi i}{3}%
},1\right) ,  \notag \\
D_{R}^{1} &\sim &\left( \mathbf{1}_{0}\mathbf{,-}1,e^{\frac{2\pi i}{3}},e^{%
\frac{2\pi i}{7}}\right) ,\hspace{1cm}D_{R}^{2}\sim \left( \mathbf{1}_{0}%
\mathbf{,-}1,e^{\frac{2\pi i}{3}},e^{\frac{\pi i}{7}}\right) ,\hspace{1cm}%
D_{R}^{3}\sim \left( \mathbf{1}_{0}\mathbf{,-}1,e^{\frac{2\pi i}{3}%
},1\right) ,  \notag \\
T_{R} &\sim &\left( \mathbf{1}_{0}\mathbf{,}1,1,1\right) ,\hspace{1cm}%
J_{R}^{1}\sim \left( \mathbf{1}_{0}\mathbf{,}1,1,e^{\frac{2\pi i}{7}}\right)
,\hspace{1cm}J_{R}^{2}\sim \left( \mathbf{1}_{0}\mathbf{,}1,1,e^{\frac{\pi i%
}{7}}\right) .
\end{eqnarray}%
Here the dimensions of the $T_{7}$ irreducible representations are specified
by the numbers in boldface. 

\quad With the aforementioned field content of our model, the relevant quark
and lepton Yukawa terms invariant under the group $\mathcal{G}$, take the
form: 
\begin{eqnarray}
-\mathcal{L}_{Y}^{\left( Q\right) } &=&y_{11}^{\left( U\right) }\overline{Q}%
_{L}^{1}\rho ^{\ast }U_{R}^{1}\frac{\sigma ^{4}}{\Lambda ^{4}}%
+y_{12}^{\left( U\right) }\overline{Q}_{L}^{1}\rho ^{\ast }U_{R}^{2}\frac{%
\sigma ^{3}}{\Lambda ^{3}}+y_{21}^{\left( U\right) }\overline{Q}_{L}^{2}\rho
^{\ast }U_{R}^{1}\frac{\sigma ^{3}}{\Lambda ^{3}}+y_{22}^{\left( U\right) }%
\overline{Q}_{L}^{2}\rho ^{\ast }U_{R}^{2}\frac{\sigma ^{2}}{\Lambda ^{2}} 
\notag \\
&&+y_{13}^{\left( U\right) }\overline{Q}_{L}^{1}\rho ^{\ast }U_{R}^{3}\frac{%
\sigma ^{2}}{\Lambda ^{2}}+y_{31}^{\left( U\right) }\overline{Q}_{L}^{3}\eta
U_{R}^{1}\frac{\sigma ^{2}}{\Lambda ^{2}}+y_{23}^{\left( U\right) }\overline{%
Q}_{L}^{2}\rho ^{\ast }U_{R}^{3}\frac{\sigma }{\Lambda }+y_{32}^{\left(
U\right) }\overline{Q}_{L}^{3}\eta U_{R}^{2}\frac{\sigma }{\Lambda }  \notag
\\
&&+y_{33}^{\left( U\right) }\overline{Q}_{L}^{3}\eta U_{R}^{3}+y^{\left(
T\right) }\overline{Q}_{L}^{3}\chi T_{R}+y_{1}^{\left( J\right) }\overline{Q}%
_{L}^{1}\chi ^{\ast }J_{R}^{1}+y_{2}^{\left( J\right) }\overline{Q}%
_{L}^{2}\chi ^{\ast }J_{R}^{2}+y_{33}^{\left( D\right) }\overline{Q}%
_{L}^{3}\rho D_{R}^{3}\frac{\tau ^{3}}{\Lambda ^{3}}  \label{Yukawaterms} \\
&&+y_{11}^{\left( D\right) }\overline{Q}_{L}^{1}\eta ^{\ast }D_{R}^{1}\frac{%
\sigma ^{4}\tau ^{3}}{\Lambda ^{7}}+y_{12}^{\left( D\right) }\overline{Q}%
_{L}^{1}\eta ^{\ast }D_{R}^{2}\frac{\sigma ^{3}\tau ^{3}}{\Lambda ^{6}}%
+y_{21}^{\left( D\right) }\overline{Q}_{L}^{2}\eta ^{\ast }D_{R}^{1}\frac{%
\sigma ^{3}\tau ^{3}}{\Lambda ^{6}}+y_{22}^{\left( D\right) }\overline{Q}%
_{L}^{2}\eta ^{\ast }D_{R}^{2}\frac{\sigma ^{2}\tau ^{3}}{\Lambda ^{5}} 
\notag \\
&&+y_{13}^{\left( D\right) }\overline{Q}_{L}^{1}\eta ^{\ast }D_{R}^{3}\frac{%
\sigma ^{2}\tau ^{3}}{\Lambda ^{5}}+y_{31}^{\left( D\right) }\overline{Q}%
_{L}^{3}\rho D_{R}^{1}\frac{\sigma ^{2}\tau ^{3}}{\Lambda ^{5}}%
+y_{23}^{\left( D\right) }\overline{Q}_{L}^{2}\eta ^{\ast }D_{R}^{3}\frac{%
\sigma \tau ^{3}}{\Lambda }+y_{32}^{\left( D\right) }\overline{Q}%
_{L}^{3}\rho D_{R}^{2}\frac{\sigma \tau ^{3}}{\Lambda }+H.c,  \notag
\end{eqnarray}
\begin{eqnarray}
-\mathcal{L}_{Y}^{\left( L\right) } &=&h_{\rho e}^{\left( L\right) }\left( 
\overline{L}_{L}\rho \xi \right) _{\mathbf{\mathbf{1}}_{0}}e_{R}\frac{\sigma
^{7}}{\Lambda ^{8}}+h_{\rho \mu }^{\left( L\right) }\left( \overline{L}%
_{L}\rho \xi \right) _{\mathbf{1}_{2}}\mu _{R}\frac{\sigma ^{4}}{\Lambda ^{5}%
}+h_{\rho \tau }^{\left( L\right) }\left( \overline{L}_{L}\rho \xi \right) _{%
\mathbf{1}_{1}}\tau _{R}\frac{\sigma ^{2}}{\Lambda ^{3}}  \notag \\
&&+h_{\chi }^{\left( L\right) }\left( \overline{L}_{L}\chi N_{R}\right) _{%
\mathbf{\mathbf{1}}_{0}}+\frac{1}{2}h_{1N}\left( \overline{N}%
_{R}N_{R}^{C}\right) _{\mathbf{3}}\xi ^{\ast }+h_{2N}\left( \overline{N}%
_{R}N_{R}^{C}\right) _{\overline{\mathbf{3}}}S  \notag \\
&&+h_{\rho }\varepsilon _{abc}\left( \overline{L}_{L}^{a}\left(
L_{L}^{C}\right) ^{b}\right) _{\mathbf{3}}(\rho^*)^{c}\frac{\zeta }{\Lambda }%
+H.c,  \label{Lylepton}
\end{eqnarray}%
where $y_{ij}^{\left( U,D\right) }$ ($i,j=1,2,3$), $h_{\rho e}^{\left(
L\right) }$, $h_{\rho \mu }^{\left( L\right) }$, $h_{\rho \tau }^{\left(
L\right) }$, $h_{\chi }^{\left( L\right) }$, $h_{1N}$, $h_{2N}$ and $h_{\rho
}$ are $\mathcal{O}(1)$ dimensionless couplings.

\quad In the following we explain the role each discrete group factors of
our model. The $T_{7}$ and $Z_{3}$ discrete groups reduce the number of the $%
SU(3)_{C}\otimes SU(3)_{L}\otimes U(1)_{X}$ model parameters. We use $T_{7}$
since it is the minimal non-Abelian discrete group having a complex triplet 
\cite{Ishimori:2010au}, where the three fermion generations can be naturally
unified. The $Z_{3}$ symmetry determines the allowed entries of the neutrino
mass matrix and forbids mixings between SM quarks and exotic quarks. The $%
Z_{2}$ symmetry is responsible for the mass hierarchy between SM up and SM
down type quarks. 
The $Z_{14}$ symmetry generates the hierarchy among charged fermion masses
and quark mixing angles that yields the observed charged fermion mass and
quark mixing pattern. 
We use $Z_{14}$ because it is the smallest lowest cyclic symmetry, that
allows to build a twelve dimensional charged lepton Yukawa term crucial to
explain the smallness of the electron mass, without tuning its corresponding
Yukawa coupling. 

\quad To get a predictive model that successfully accounts for fermion
masses and mixings, we assume that the $SU(3)_{L}$ singlet scalars have the
following VEV pattern: 
\begin{equation}
\left\langle \sigma \right\rangle =v_{\sigma }e^{i\phi },\hspace{0.5cm}%
\left\langle \tau \right\rangle =v_{\tau },\hspace{0.5cm}\left\langle \xi
\right\rangle =\frac{v_{\xi }}{\sqrt{3}}\left( 1,1,1\right) ,\hspace{0.5cm}%
\left\langle \zeta \right\rangle =\frac{v_{\zeta }}{\sqrt{2}}\left(
1,0,1\right) ,\hspace{0.5cm}\left\langle S\right\rangle =\frac{v_{S}}{\sqrt{3%
}}\left( 1,1,-1\right) .  \label{VEV}
\end{equation}
which we have checked to be consistent with the scalar potential
minimization equations. 

\quad Besides that, the $SU(3)_{L}$ scalar singlets are assumed to acquire
vacuum expectation values at a very high energy $\Lambda _{int}\gg v_{\chi
}\approx \mathcal{O}(1)$ TeV, 
excepting $\zeta _{j}$ ($j=1,2,3$), whose vacuum expectation value is much
lower than the scale of electroweak symmetry breaking $v=246$ GeV. Let's
note that at the scale $\Lambda _{int}$, the $SU(3)_{C}\otimes
SU(3)_{L}\otimes U(1)_{X}\otimes T_{7}\otimes Z_{2}\otimes Z_{3}\otimes
Z_{14}$ symmetry is broken to $SU(3)_{C}\otimes SU(3)_{L}\otimes U(1)_{X}$
by the vacuum expectation values of the $SU(3)_{L}$ singlet scalar fields $%
\xi _{j}$, $S_{j}$, $\sigma $ and $\tau $. 

\quad Considering that the charged fermion mass and quark mixing pattern
arises from the $Z_{2}\otimes Z_{3}\otimes Z_{14}$ symmetry breaking, we set
the VEVs of the $SU(3)_{L}$ singlet scalars $S$, $\xi $, $\sigma $ and $\tau 
$, as follows: 
\begin{equation}
v_{S}\sim v_{\xi }=v_{\sigma }=v_{\tau }=\Lambda _{int}=\lambda \Lambda ,
\label{VEVsinglets}
\end{equation}%
being $\lambda =0.225$ one of the Wolfenstein parameters and $\Lambda $ our
model cutoff. Consequently, the VEVs of the scalars in our model have the
following hierarchy: 
\begin{equation}
v_{\zeta }<<v_{\rho }\sim v_{\eta }\sim v<<v_{\chi }<<\Lambda _{int}.
\end{equation}%
%
%
%
%
%
%

%
%

\section{Lepton masses and mixings}

\label{leptonmassesandmixing} From Eq. (\ref{Lylepton}), it follows that
the mass matrix for charged leptons is \cite{Hernandez:2015cra}: 
\begin{equation}
M_{l}=V_{lL}^{\dag }P_{l}diag\left( m_{e},m_{\mu },m_{\tau }\right) ,\hspace{%
0.5cm}V_{lL}=\frac{1}{\sqrt{3}}\left( 
\begin{array}{ccc}
1 & 1 & 1 \\ 
1 & \omega & \omega ^{2} \\ 
1 & \omega ^{2} & \omega%
\end{array}%
\right) ,\hspace{0.5cm}P_{l}=\left( 
\begin{array}{ccc}
e^{7i\phi } & 0 & 0 \\ 
0 & e^{4i\phi } & 0 \\ 
0 & 0 & e^{2i\phi }%
\end{array}%
\right) ,\hspace{0.5cm}\omega =e^{\frac{2\pi i}{3}},
\end{equation}%
%
%
%
%
%
%
%
%
%
%
%
%
%
%
%
%
%
%
%
%
%
%
%
%
%
%
%
%
%
%
%
where the charged lepton masses read:
\begin{equation}
m_{e}=h_{\rho e}^{\left( L\right) }\lambda ^{8}\frac{v_{\rho }}{\sqrt{2}},%
\hspace{1cm}m_{\mu }=h_{\rho \mu }^{\left( L\right) }\lambda ^{5}\frac{%
v_{\rho }}{\sqrt{2}},\hspace{1cm}m_{\tau }=h_{\rho \tau }^{\left( L\right)
}\lambda ^{3}\frac{v_{\rho }}{\sqrt{2}}.  \label{leptonmasses}
\end{equation}%
Taking into account that $v_{\rho }\approx v=246$ GeV, it follows that the
charged lepton masses are related with the electroweak symmetry breaking
scale by their scalings with powers of the Wolfenstein parameter $\lambda
=0.225$, with $\mathcal{O}(1)$ coefficients. 

\quad The neutrino mass matrix is given by \cite{Hernandez:2015cra}: 
\begin{eqnarray}
M_{\nu } &=&\left( 
\begin{array}{ccc}
0_{3\times 3} & M_{D} & 0_{3\times 3} \\ 
M_{D}^{T} & 0_{3\times 3} & M_{\chi } \\ 
0_{3\times 3} & M_{\chi }^{T} & M_{R}%
\end{array}%
\right) ,\hspace{0.7cm}M_{D}=\frac{h_{\rho }v_{\rho }v_{\zeta }}{2\Lambda }%
\left( 
\begin{array}{ccc}
0 & 1 & 0 \\ 
-1 & 0 & -1 \\ 
0 & 1 & 0%
\end{array}%
\right) ,\hspace{0.7cm}M_{\chi }=h_{\chi }^{\left( L\right) }\frac{v_{\chi }%
}{\sqrt{2}}\left( 
\begin{array}{ccc}
1 & 0 & 0 \\ 
0 & 1 & 0 \\ 
0 & 0 & 1%
\end{array}%
\right) ,  \notag \\
M_{R} &=&h_{1N}\frac{v_{\xi }}{\sqrt{3}}\left( 
\begin{array}{ccc}
1 & -x & x \\ 
-x & 1 & x \\ 
x & x & 1%
\end{array}%
\right) ,\hspace{0.7cm}x=\frac{h_{2N}v_{S}}{h_{1N}v_{\xi }}.
\end{eqnarray}

Since the $SU(3)_{L}$ singlet scalars having Yukawa interactions with the
right handed Majorana neutrinos acquire VEVs at very high scale, these
Majorana neutrinos are very heavy, so that the active neutrinos get small
masses via a double seesaw mechanism.

The neutrino mass matrices for the physical states are \cite{Catano:2012kw}: 
\begin{equation}
M_{\nu }^{\left( 1\right) }=M_{D}\left( M_{\chi }^{T}\right)
^{-1}M_{R}M_{\chi }^{-1}M_{D}^{T},\hspace{0.5cm}\hspace{0.5cm}M_{\nu
}^{\left( 2\right) }=-M_{\chi }M_{R}^{-1}M_{\chi }^{T},\hspace{0.5cm}\hspace{%
0.5cm}M_{\nu }^{\left( 3\right) }=M_{R},  \label{Mnu1}
\end{equation}%
being $M_{\nu }^{\left( 1\right) }$ the mass matrix for light active
neutrinos, while $M_{\nu }^{\left( 2\right) }$ and $M_{\nu }^{\left(
3\right) }$ are the heavy and very heavy sterile neutrino mass matrices,
respectively. Consequently, the double seesaw mechanism gives rise to light
active neutrinos as well as to heavy and very heavy sterile neutrinos.

\quad Using Eq. (\ref{Mnu1}), we find the following mass matrix for light
active neutrinos \cite{Hernandez:2015cra}: 
\begin{equation}
M_{\nu }^{\left( 1\right) }=\left( 
\begin{array}{ccc}
A & 0 & A \\ 
0 & B & 0 \\ 
A & 0 & A%
\end{array}%
\right) ,\hspace{1cm}A=\frac{h_{1N}h_{\rho }^{2}v_{\rho }^{2}v_{\zeta
}^{2}v_{\xi }}{2\sqrt{3}h_{\chi }^{\left( L\right) }v_{\chi }^{2}\Lambda ^{2}%
},\hspace{1cm}B=\frac{h_{\rho }^{2}v_{\rho }^{2}v_{\zeta }^{2}}{\sqrt{3}%
h_{\chi }^{\left( L\right) }v_{\chi }^{2}\Lambda ^{2}}\left( h_{1N}v_{\xi
}+h_{2N}v_{S}\right) .  \label{Mnu}
\end{equation}%
From Eq. (\ref{Mnu}) it follows that the light active neutrino mass matrix 
only depends on two effective parameters: $A$ and $B$, which determine the
neutrino mass squared splittings. 
Let's note that $A$ and $B$ are supressed by their scaling with inverse
powers of the high energy cutoff $\Lambda $. Furthermore, we have that the
smallness of the active neutrino masses arises from their scaling with
inverse powers of the high energy cutoff $\Lambda $ as well as from their
quadratic dependence on the very small VEV of the $Z_{2}\otimes Z_{3}\otimes
Z_{14}$ neutral, $SU(3)_{L}$ singlet and $T_{7}$ antitriplet scalar field $%
\zeta $. From Eq. (\ref{Mnu}) and the relations $v_{\xi }=\lambda \Lambda $, 
$v_{\rho }\sim 100$ GeV, $v_{\chi }\sim 1$ TeV,\ we get that the mass scale
for the light active neutrinos satisfies $m_{\nu }\sim 10^{-3}\frac{v_{\zeta
}^{2}}{\Lambda }$. Consequently, setting $v_{\zeta }=1$ GeV, we find for the
cutoff of our model the estimate 
\begin{equation}
\Lambda \sim 10^{5}\text{ TeV},  \label{cutoff}
\end{equation}%
which is of the same order of magnitude of the cutoff of our $S_{3}$ lepton
flavor $331$ model \cite{Hernandez:2014lpa}. Consequenty, we find that the
heavy and very heavy sterile neutrinos have masses at the $\sim $ MeV and $%
\sim $ TeV scales, respectively. Then the MeV sterile neutrinos correspond
to dark matter candidates. Furthermore, we assume that the lightest of the
very heavy sterile neutrinos, i.e., $\xi _{1L}^{\left( 3\right) }$ has a TeV
scale mass and thus corresponds to a candidate for detection at the LHC.

Moreover, we find that the mass matrix $M_{\nu }^{\left( 1\right) }$ for
light active neutrinos is diagonalized by a rotation matrix $V_{\nu }$, as
follows: 
\begin{equation}
V_{\nu }^{T}M_{\nu }^{\left( 1\right) }V_{\nu }=\left( 
\begin{array}{ccc}
m_{1} & 0 & 0 \\ 
0 & m_{2} & 0 \\ 
0 & 0 & m_{3}%
\end{array}%
\right) ,\hspace{0.5cm}\mbox{with}\hspace{0.5cm}V_{\nu }=\left( 
\begin{array}{ccc}
\cos \theta & 0 & \sin \theta \\ 
0 & 1 & 0 \\ 
-\sin \theta & 0 & \cos \theta%
\end{array}%
\right) ,\hspace{0.5cm}\theta =\pm \frac{\pi }{4},  \label{Vnu}
\end{equation}%
where $\theta =+\pi /4$ and $\theta =-\pi /4$ correspond to normal (NH) and
inverted (IH) mass hierarchies, respectively. The masses for the light
active neutrinos, in the cases of normal (NH) and inverted (IH) mass
hierarchies, read: 
\begin{eqnarray}
\mbox{NH} &:&\theta =+\frac{\pi }{4}:\hspace{10mm}m_{\nu _{1}}=0,\hspace{10mm%
}m_{\nu _{2}}=B,\hspace{10mm}m_{\nu _{3}}=2A,  \label{mass-spectrum-Inverted}
\\[0.12in]
\mbox{IH} &:&\theta =-\frac{\pi }{4}:\hspace{10mm}m_{\nu _{1}}=2A,\hspace{8mm%
}m_{\nu _{2}}=B,\hspace{10mm}m_{\nu _{3}}=0.  \label{mass-spectrum-Normal}
\end{eqnarray}%
%
%
%
%
%
%
\quad Furthermore, we find that the lepton mixing angles are given by: 
\begin{equation}
\sin ^{2}\theta _{12}=\frac{1}{2\mp \cos 5\phi },\hspace{0.5cm}\sin
^{2}\theta _{13}=\frac{1}{3}(1\pm \cos 5\phi ),\hspace{0.5cm}\sin ^{2}\theta
_{23}=\frac{1}{2}\pm \frac{\sqrt{3}\sin 5\phi}{4\mp 2\cos 5\phi }.
\label{theta-ij}
\end{equation}%
Then, it follows that the limit $\phi =0$ and $\phi =\pi $ for the inverted
and normal neutrino mass hierarchies, respectively, correspond to the
tribimaximal mixing, which predicts a vanishing reactor mixing angle. 
Let's note that the mixing angles for the lepton sector only depend on a
single parameter ($\phi $), while the neutrino mass squared splittings are
controlled by two parameters, i.e., $A$ and $B$. Furthermore, from the
relation $\theta =\pm \frac{\pi }{4}$, we predict $J=0$ and $\delta =0$,
which implies that our model predicts a vanishing leptonic Dirac CP
violating phase.%

\quad 
The parameters $A$ and $B$ for the normal (NH) and inverted (IH) neutrino
mass hierarchies read: 
\begin{eqnarray}
&&\mbox{NH}:\ m_{\nu _{1}}=0,\ \ \ m_{\nu _{2}}=B=\sqrt{\Delta m_{21}^{2}}%
\approx 9\mbox{meV},\ \ \ m_{\nu _{3}}=2A=\sqrt{\Delta m_{31}^{2}}\approx 50%
\mbox{meV};  \label{AB-Delta-IH} \\[0.12in]
&&\mbox{IH}\hspace{2mm}:\ m_{\nu _{2}}=B=\sqrt{\Delta m_{21}^{2}+\Delta
m_{13}^{2}}\approx 50\mbox{meV},\ \ \ m_{\nu _{1}}=2A=\sqrt{\Delta
m_{13}^{2}}\approx 49\mbox{meV},\ \ \  m_{\nu _{3}}=0,\nonumber  \label{AB-Delta-NH}
\end{eqnarray}%
which follows from Eqs. (\ref{mass-spectrum-Normal}), (\ref%
{mass-spectrum-Inverted}) and the definition $\Delta
m_{ij}^{2}=m_{i}^{2}-m_{j}^{2}$. We take the best fit values of $\Delta
m_{ij}^{2}$ from Tables \ref{NH} and \ref{IH} for the normal and inverted
neutrino mass hierarchies, respectively.

\quad 
To reproduce the experimental values of the leptonic mixing parameters $\sin
^{2}\theta _{ij}$ given in Tables \ref{NH}, \ref{IH}, we vary the $\phi $
parameter, finding the following result: 
\begin{eqnarray}
&&\mbox{NH}\ :\ \phi =0.576\,\pi ,\ \ \ \sin ^{2}\theta _{12}\approx 0.34,\
\ \ \sin ^{2}\theta _{23}\approx 0.61,\ \ \ \sin ^{2}\theta _{13}\approx
0.0232;  \label{parameter-fit-IH} \\[0.12in]
&&\mbox{IH}\hspace{2.5mm}:\ \phi =\ \ 0.376\pi ,\ \ \ \ \ \sin ^{2}\theta
_{12}\approx 0.34,\ \ \ \sin ^{2}\theta _{23}\approx 0.61,\ \ \ \ \,\sin
^{2}\theta _{13}\approx 0.0238.  \label{parameter-fit-NH}
\end{eqnarray}

\quad Consequently, we find 
that $\sin ^{2}\theta _{13}$ is in excellent agreement with the experimental
data, for both normal and inverted neutrino mass hierarchies, whereas $\sin
^{2}\theta _{12}$ and $\sin ^{2}\theta _{23}$\ stay in the experimentally
allowed $2\sigma $ 
range. Thus, our predictions for the neutrino mass squared splittings and
leptonic mixing parameters, 
are in very good agreement with the experimental data on neutrino
oscillations, for both normal and inverted mass hierarchies.
\begin{table}[tbh]
\begin{tabular}{|c|c|c|c|c|c|}
\hline
Parameter & $\Delta m_{21}^{2}$($10^{-5}$eV$^2$) & $\Delta m_{31}^{2}$($%
10^{-3}$eV$^2$) & $\left( \sin ^{2}\theta _{12}\right) _{\exp }$ & $\left(
\sin ^{2}\theta _{23}\right) _{\exp }$ & $\left( \sin ^{2}\theta
_{13}\right) _{\exp }$ \\ \hline
Best fit & $7.60$ & $2.48$ & $0.323$ & $0.567$ & $0.0234$ \\ \hline
$1\sigma $ range & $7.42-7.79$ & $2.41-2.53$ & $0.307-0.339$ & $0.439-0.599$
& $0.0214-0.0254$ \\ \hline
$2\sigma $ range & $7.26-7.99$ & $2.35-2.59$ & $0.292-0.357$ & $0.413-0.623$
& $0.0195-0.0274$ \\ \hline
$3\sigma $ range & $7.11-8.11$ & $2.30-2.65$ & $0.278-0.375$ & $0.392-0.643$
& $0.0183-0.0297$ \\ \hline
\end{tabular}%
\caption{Range for experimental values of neutrino mass squared splittings and leptonic mixing parameters, taken from Ref. \cite{Forero:2014bxa}, for the case of normal hierarchy.}
\label{NH}
\end{table}
\begin{table}[tbh]
\begin{tabular}{|c|c|c|c|c|c|}
\hline
Parameter & $\Delta m_{21}^{2}$($10^{-5}$eV$^{2}$) & $\Delta m_{13}^{2}$($%
10^{-3}$eV$^{2}$) & $\left( \sin ^{2}\theta _{12}\right) _{\exp }$ & $\left(
\sin ^{2}\theta _{23}\right) _{\exp }$ & $\left( \sin ^{2}\theta
_{13}\right) _{\exp }$ \\ \hline
Best fit & $7.60$ & $2.38$ & $0.323$ & $0.573$ & $0.0240$ \\ \hline
$1\sigma $ range & $7.42-7.79$ & $2.32-2.43$ & $0.307-0.339$ & $0.530-0.598$
& $0.0221-0.0259$ \\ \hline
$2\sigma $ range & $7.26-7.99$ & $2.26-2.48$ & $0.292-0.357$ & $0.432-0.621$
& $0.0202-0.0278$ \\ \hline
$3\sigma $ range & $7.11-8.11$ & $2.20-2.54$ & $0.278-0.375$ & $0.403-0.640$
& $0.0183-0.0297$ \\ \hline
\end{tabular}%
\caption{Range for experimental values of neutrino mass squared splittings and leptonic mixing parameters, taken from Ref. \cite{Forero:2014bxa}, for the case of inverted hierarchy.}
\label{IH}
\end{table}

\section{Quark masses and mixing.}

\label{quarkmassesandmixing} From Eq. (\ref{Yukawaterms}), it follows that the SM quark mass matrices have the form \cite{Hernandez:2015cra}: 
\begin{equation}
M_{U}=\left( 
\begin{array}{ccc}
a_{11}^{\left( U\right) }\lambda ^{4} & a_{12}^{\left( U\right) }\lambda ^{3}
& a_{13}^{\left( U\right) }\lambda ^{2} \\ 
a_{21}^{\left( U\right) }\lambda ^{3} & a_{22}^{\left( U\right) }\lambda ^{2}
& a_{23}^{\left( U\right) }\lambda \\ 
a_{31}^{\left( U\right) }\lambda ^{2} & a_{32}^{\left( U\right) }\lambda & 
a_{33}^{\left( U\right) }%
\end{array}%
\right) \frac{v}{\sqrt{2}},\hspace{1cm}\hspace{1cm}M_{D}=\left( 
\begin{array}{ccc}
a_{11}^{\left( D\right) }\lambda ^{7} & a_{12}^{\left( D\right) }\lambda ^{6}
& a_{13}^{\left( D\right) }\lambda ^{5} \\ 
a_{21}^{\left( D\right) }\lambda ^{6} & a_{22}^{\left( D\right) }\lambda ^{5}
& a_{23}^{\left( D\right) }\lambda ^{4} \\ 
a_{31}^{\left( D\right) }\lambda ^{5} & a_{32}^{\left( D\right) }\lambda ^{4}
& a_{33}^{\left( U\right) }\lambda ^{3}%
\end{array}%
\right) \frac{v}{\sqrt{2}},  \label{Mq}
\end{equation}%
where 
$a_{ij}^{\left( U,D\right) }$ ($i,j=1,2,3$) are $\mathcal{O}(1)$ parameters.
Furthermore, the exotic quark masses read:\vspace{-0.2cm}
\begin{equation}
m_{T}=y^{\left( T\right) }\frac{v_{\chi }}{\sqrt{2}},\hspace{1cm}%
m_{J^{1}}=y_{1}^{\left( J\right) }\frac{v_{\chi }}{\sqrt{2}}=\frac{%
y_{1}^{\left( J\right) }}{y^{\left( T\right) }}m_{T},\hspace{1cm}%
m_{J^{2}}=y_{2}^{\left( J\right) }\frac{v_{\chi }}{\sqrt{2}}=\frac{%
y_{2}^{\left( J\right) }}{y^{\left( T\right) }}m_{T}.  \label{mexotics}
\end{equation}
Since the charged fermion mass and quark mixing pattern arises from the
breaking of the $Z_{2}\otimes Z_{3}\otimes Z_{14}$ discrete group, we
asssume an approximate universality in the dimensionless SM quark Yukawa
couplings, as follows: 
\begin{eqnarray}
a_{11}^{\left( U\right) } &=&a_{1}^{\left( U\right) }e^{i\phi _{1}},\hspace{%
0.5cm}a_{22}^{\left( U\right) }=a_{2}^{\left( U\right) },\hspace{0.5cm}%
a_{33}^{\left( U\right) }=a_{3}^{\left( U\right) },  \label{aQ} \\
a_{12}^{\left( U\right) } &=&a_{1}^{\left( U\right) }\left( 1-\frac{\lambda
^{2}}{2}\right) ^{-\frac{3}{2}}e^{i\phi _{2}},\hspace{0.5cm}a_{13}^{\left(
U\right) }=a_{2}^{\left( U\right) }\left( 1-\frac{\lambda ^{2}}{2}\right) ^{-%
\frac{3}{2}}e^{i\phi _{2}},\hspace{0.5cm}a_{23}^{\left( U\right)
}=\left\vert a_{13}^{\left( U\right) }\right\vert \left( 1-\frac{\lambda ^{2}%
}{2}\right) ^{-\frac{3}{2}},  \notag \\
a_{11}^{\left( D\right) } &=&a_{22}^{\left( D\right) }\left( 1-\frac{\lambda
^{2}}{2}\right) ^{-2},\hspace{0.5cm}a_{23}^{\left( D\right) }=a_{33}^{\left(
D\right) }\left( 1-\frac{\lambda ^{2}}{2}\right) ^{-\frac{1}{2}},\hspace{%
0.5cm}a_{ij}^{\left( U,D\right) }=a_{ji}^{\left( U,D\right) },\hspace{0.5cm}%
i,j=1,2,3,  \notag
\end{eqnarray}
To generate the up, down, strange and charm quark masses, the universality in the quark Yukawa couplings has to be broken. 
Besides that, for simplicity, we assume that the complex phase responsible
for CP violation in the quark sector only arises from up type quark Yukawa
terms, as indicated by Eq. (\ref{aQ}). In addition, for the sake of
simplicity, we fix $a_{3}^{\left( U\right) }=1$, which is suggested by
naturalness arguments. Let's recall that the quark sector has 10 effective
parameters, i.e, $\lambda $, $a_{3}^{\left( U\right) }$, $a_{1}^{\left(
U\right) }$, $a_{2}^{\left( U\right) }$, $a_{22}^{\left( D\right) }$, $%
a_{12}^{\left( D\right) }$, $a_{13}^{\left( D\right) }$, $a_{33}^{\left(
D\right) }$ and the phases $\phi _{1}$ and $\phi _{2}$ to describe the quark
mass and mixing pattern, which is determined by 10 observables.
Nevertheless, not all these effective parameters are free since the
parameters $\lambda $ and $a_{3}^{\left( U\right) }$ are fixed while the
remaining 8 parameters are adjusted to reproduce the physical observables in
the quark sector, i.e., 6 quark masses and 4 quark mixing parameters. The
results shown in Table \ref{Observables} correspond to the following values: 
\begin{eqnarray}
a_{1}^{\left( U\right) } &\simeq &0.64,\hspace{1cm}a_{2}^{\left( U\right)
}\simeq 0.77,\hspace{1cm}\phi _{1}\simeq -9.03^{\circ },\hspace{1cm}\phi
_{2}\simeq -4.53^{\circ },  \notag \\
a_{22}^{\left( D\right) } &\simeq &2.03,\hspace{1cm}a_{12}^{\left( D\right)
}\simeq 1.75,\hspace{1cm}a_{13}^{\left( D\right) }\simeq 1.15,\hspace{1cm}%
a_{33}^{\left( D\right) }\simeq 1.40.
\end{eqnarray}
\vspace{-0.5cm}
\begin{table}[tbh]
\begin{center}
\begin{tabular}{c|l|l}
\hline\hline
Observable & Model value & Experimental value \\ \hline
$m_{u}(MeV)$ & \quad $1.59$ & \quad $1.45_{-0.45}^{+0.56}$ \\ \hline
$m_{c}(MeV)$ & \quad $673$ & \quad $635\pm 86$ \\ \hline
$m_{t}(GeV)$ & \quad $180$ & \quad $172.1\pm 0.6\pm 0.9$ \\ \hline
$m_{d}(MeV)$ & \quad $2.9$ & \quad $2.9_{-0.4}^{+0.5}$ \\ \hline
$m_{s}(MeV)$ & \quad $59.7$ & \quad $57.7_{-15.7}^{+16.8}$ \\ \hline
$m_{b}(GeV)$ & \quad $2.98$ & \quad $2.82_{-0.04}^{+0.09}$ \\ \hline
$\bigl|V_{ud}\bigr|$ & \quad $0.975$ & \quad $0.97427\pm 0.00015$ \\ \hline
$\bigl|V_{us}\bigr|$ & \quad $0.224$ & \quad $0.22534\pm 0.00065$ \\ \hline
$\bigl|V_{ub}\bigr|$ & \quad $0.0036$ & \quad $0.00351_{-0.00014}^{+0.00015}$
\\ \hline
$\bigl|V_{cd}\bigr|$ & \quad $0.224$ & \quad $0.22520\pm 0.00065$ \\ \hline
$\bigl|V_{cs}\bigr|$ & \quad $0.9736$ & \quad $0.97344\pm 0.00016$ \\ \hline
$\bigl|V_{cb}\bigr|$ & \quad $0.0433$ & \quad $0.0412_{-0.0005}^{+0.0011}$
\\ \hline
$\bigl|V_{td}\bigr|$ & \quad $0.00853$ & \quad $%
0.00867_{-0.00031}^{+0.00029} $ \\ \hline
$\bigl|V_{ts}\bigr|$ & \quad $0.0426$ & \quad $0.0404_{-0.0005}^{+0.0011}$
\\ \hline
$\bigl|V_{tb}\bigr|$ & \quad $0.999057$ & \quad $%
0.999146_{-0.000046}^{+0.000021}$ \\ \hline
$J$ & \quad $2.98\times 10^{-5}$ & \quad $(2.96_{-0.16}^{+0.20})\times
10^{-5}$ \\ \hline
$\delta $ & \quad $61^{\circ }$ & \quad $68^{\circ }$ \\ \hline\hline
\end{tabular}%
\end{center}
\caption{Model and experimental values of the quark masses and CKM
parameters.}
\label{Observables}
\end{table}

The obtained and experimental values of the quark masses, CKM matrix
elements, Jarlskog invariant $J$ and CP violating phase $\delta $ are
reported in Table \ref{Observables}. We use the experimental values of the
quark masses at the $M_{Z}$ scale, from Ref. (\cite{Bora:2012tx}), whereas
we use the experimental values of the CKM parameters from Ref. \cite%
{Agashe:2014kda}. The obtained values of the quark masses and CKM parameters
are in excellent agreement with the experimental data, as indicated by Table %
\ref{Observables}. 

\section{Conclusions}

\label{conclusions}
We presented an extension of the minimal $331$ model with $\beta
=-\frac{1}{\sqrt{3}}$, based on the extended $SU(3)_{C}\otimes
SU(3)_{L}\otimes U(1)_{X}\otimes T_{7}\otimes Z_{2}\otimes Z_{3}\otimes
Z_{14}$ symmetry, compatible with the experimental data on fermion masses
and mixing. The $T_{7}$ and $Z_{3}$ symmetries reduce the number of model
parameters. In particular, the $Z_{3}$ symmetry determines the allowed
entries of the neutrino mass matrix and decouples the SM quarks from the
exotic quarks. The $Z_{2}$ symmetry generates the hierarchy between SM up
and SM down type quark masses. We assumed that the $SU(3)_{L}$ scalar
singlets having Yukawa interactions with the right handed Majorana neutrinos
acquire VEVs at very high scale, then providing very large masses to these
Majorana neutrinos, and thus giving rise to a double seesaw mechanism of
active neutrino masses. 
Consequently, the neutrino spectrum includes very light active neutrinos as
well as heavy and very heavy sterile neutrinos. We find that the heavy and
very heavy sterile neutrinos have masses at the $\sim $ MeV and $\sim $ TeV
scales, respectively. Thus, the MeV scale sterile neutrinos of our model
correspond to dark matter candidates. The smallness of the active neutrino
masses is attributed to their scaling with inverse powers of the high energy
cutoff $\Lambda \sim 10^{5}$ TeV as well as well as by their quadratic
dependence on the very small VEV of the $Z_{2}\otimes Z_{3}\otimes Z_{14}$
neutral, $SU(3)_{L}$ singlet and $T_{7}$ antitriplet scalar field $\zeta $.
The observed hierarchy of charged fermion masses and quark mixing matrix
elements arises from the breaking of the $Z_{2}\otimes Z_{3}\otimes Z_{14}$
discrete group at a very high energy. Furthermore, our model predicts a
vanishing leptonic Dirac CP violating phase.

\section*{Acknowledgments}

A.E.C.H was supported by Fondecyt (Chile), Grant No. 11130115 and by DGIP
internal Grant No. 111458. R.M. was supported by COLCIENCIAS. A. E. C. H thanks the organizers of Planck 2015 for inviting him to present this talk.


\end{document}